\documentclass{emulateapj}
\begin{document}



\title{A Scalable Hybrid FPGA/GPU FX Correlator}

\author{J. Kocz$^1$, L.J. Greenhill$^1$, B.R. Barsdell$^{1}$, G. Bernardi$^{1,2,3}$,
  A. Jameson$^{4}$, M.A. Clark$^{1,5}$, J. Craig$^{6}$, D. Price$^{1}$, G.B. Taylor$^{6}$, F. Schinzel$^{6}$,
D. Werthimer$^{7}$}

\address{
$^1$Harvard-Smithsonian Center for Astrophysics, 60 Garden Street,
Cambridge, Massachusetts, 02138, USA\\
$^2$SKA SA, 3rd Floor, The Park, Park Road, Pinelands, 7405, South Africa\\
$^3$Department of Physics and Electronics, Rhodes University, PO Box
94, Grahamstown, 6140, South Africa\\
$^4$Centre for Astrophysics and Supercomputing, Swinburne University of Technology, 1 Alfred Street, Hawthorn, Victoria, 3122, Australia \\
$^5$NVIDIA Corporation, 2701 San Tomas Expressway, Santa Clara, CA 95050, USA\\
$^6$ Department of Physics and Astronomy, University of New Mexico, Albuquerque, NM 87131, USA\\
$^7$Space Sciences Lab, University of California, Berkeley, CA 94720, USA 
}




\begin{abstract}
Radio astronomical imaging arrays comprising large numbers of antennas, O(10$^2$-10$^3$) have posed a signal processing challenge because of the required O(N$^2$) cross correlation of signals from each antenna and requisite signal routing. This motivated the implementation of a Packetized Correlator architecture that applies Field Programmable Gate Arrays (FPGAs) to the O(N) ``F-stage'' transforming time domain to frequency domain data, and  Graphics Processing Units (GPUs) to the O(N$^2$) ``X-stage'' performing an outer product among spectra for each antenna.  The design is readily scalable to at least O(10$^3$) antennas. Fringes, visibility amplitudes and sky image results obtained during field testing are presented.
\end{abstract}


\section{Introduction}

Cross-correlation of time-series signals from antennas in radio astronomical arrays scales quadratically with the number of antennas or phased elements and linearly with bandwidth.  A new generation of full cross-correlation low-frequency arrays, motivated in part by the science of 21~cm cosmology \citep{furlanetto06}, is notable for the numbers of antennas employed, which may number in the hundreds.  Examples include the Long Wavelength Array \citep[LWA;][]{taylor2012} and the Precision Array for Probing the Epoch of Reionization \citep[PAPER;][]{parsons2010}.

Correlator designs that rely on Application Specific Integrated Circuits (ASICs) and Field Programmable Gate Arrays (FPGAs) are common \citep{perley2009}.  Systems have also been implemented in software, running on general purpose CPUs \citep{deller2007}.  Though ASIC and FPGA-based designs are efficient in terms of computation and signal transport, development efforts may have long lead times and require specialized digital engineering, and upgrades in capability without extensive re-engineering may be impractical.  Moreover, in both cases implementations do not depend on off-the-shelf mass manufactured hardware, use of which may reduce cost.  In contrast, CPU-based correlation depends on general purpose languages, libraries, and hardware.  It leverages investment by computer  and computational science communities in algorithms and optimizations, as well as the economies of scale in manufacturing of hardware.   The approach is inherently flexible, but the computing density of general purpose CPUs limits application to small arrays.

Graphics Processing Units (GPUs) are massively parallel computation engines that are well matched to the cross multiplication operation of the ``FX'' correlation algorithm \citep{tms01} and the high-speed movement of data that is required for large arrays, i.e., for a large number of correlator signal inputs. The operation may be phrased as an outer product of two vectors that contain data for each antenna for a given time and frequency. The problem is readily parallelized and on GPUs achieves high arithmetic intensity (number of operations per byte moved).  Although use of floating point arithmetic is intrinsically less efficient than the fixed point arithmetic typically executed by ASICs and FPGAs, high computing density, parallel architecture, high-speed memory transfer, and code reduction afforded by a compiler make GPUs strongly competitive for large arrays \citep[e.g.,][]{clark2012}.

The limitations of current correlator systems motivated the development of a new ``hybrid''\footnote{Originally the term ``hybrid'' described correlators that use analog filtering and digital frequency analysis. Such systems are now rare, and the term is repurposed here to refer to an all-digital design that mixes platforms, harnessing fixed and floating point arithmetic.} design where the F and X stages of correlation are implemented using FPGAs and GPUs respectively. The design leverages the modular architecture introduced by \citet{parsons2008}, but GPU servers are substituted for FPGA platforms and applied to the cross multiplication stage. For large N correlator systems, adoption of GPUs for execution of cross multiplication can provide an initial cost advantage \citep[e.g.,][]{filiba2013}, as well as providing the flexibility and rapid development time advantages of a software correlator. Achieving high compute utilization in high-speed stream Fourier processing of data is comparatively more difficult than in cross multiplication, due to the lower arithmetic intensity of the calculation. As a result of this FPGA platforms continue to be used to execute the synchronous digital sampling of time-series data at the correlator input and the Fourier transform to frequency space that is intrinsic to the F-stage.  However, use of GPUs throughout the processing path may also be practical for some applications in which other systems provide synchronous digital sampling. 

Section \ref{sec:corrDesign} describes the three main correlator computing layers  (FPGA, CPU, and GPU). Section \ref{sec:scale} discusses the scalability of the design. Section \ref{sec:results} introduces the results of an August 2012 field deployment, and discussion of further development options appears in Section \ref{sec:conclusions}. 

\section{Correlator Architecture}
\label{sec:corrDesign}
\subsection{Hardware}

\begin{figure*}
\includegraphics[width=16cm]{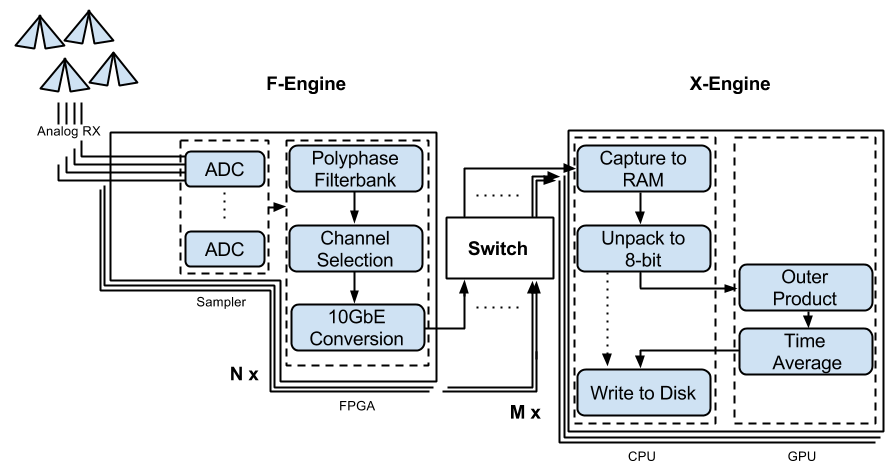}
\caption{Correlator data path.Signals are sampled by ADCs that are directly connected to the FPGA platform. Digitized time domain data are transformed to the frequency domain. A selected sub-band is formatted into ethernet packets for transmission via a network switch to a GPU server.  The data are captured to RAM, unpacked from 4 to 8-bits (complex), and copied to the GPUs for cross multiplication. The data are averaged, transferred back to server RAM and written to disk. Direct transfer to disk of spectra is enabled when sampling at the Nyquist rate is needed (dotted line).\label{fig:daq}}
\end{figure*}

The hybrid architecture can be divided into four sections (figure\,\ref{fig:daq}).  First, the F-stage implemented on FPGA hardware accepts synchronous input from digital baseband samplers,  transforms time to frequency domain data at the Nyquist rate, and formats data into Ethernet packets.  A platform such as the Reconfigurable Open Architecture Computer Hardware (ROACH) or its successor, ROACH2\footnote{http://casper.berkeley.edu/wiki/hardware}, which consists principally of an FPGA chip and the facility to connect the requisite analog to digital converter (ADC) outputs and multiple Ethernet interfaces, can be used.  Second, the network layer initiates a corner turn or transpose operation while conducting data from the F to the X stage  (see Section\,\ref{ss:ct}). This can be realized with an off-the-shelf Layer-2 switch.  Third, at entry to the X stage, packetized data capture and ordering (completing the corner turn in preparation for cross-multiplication) is executed in a CPU layer.  This can use a general purpose high-throughput server and network interface.  Fourth, cross multiplication and integration of products for each frequency channel and time sample is implemented in the GPU layer.  Averaging of time samples in the GPU layer reduces the volume of data that must be transferred back to the CPU layer following the O(N$^2$) cross multiplication.  The selection of GPU is chiefly governed by floating point computing capacity (Flop\,s$^{-1}$).  Device memory capacity is a secondary consideration, effectively reduced to O(N) needed to store data prior to correlation, due to savings afforded by the time averaging of the data afterward.  F-stage processing is parallelized by antenna input.  X-stage processing is parallelized by frequency.  Each GPU processes a sub-band extracted from the output to the F-stage.   (The number of sub-bands and thus the number of frequency channels per GPU depends on computational capacity-- completion of the O(N$^2$) calculations in a sufficiently short time for real-time processing).

In contrast to an ASIC or FPGA correlator architecture, the hybrid architecture combines synchronous and asynchronous elements, which is effective provided there is adequate buffering and the time order of data can be maintained. Each of the `F' stages are synchronized by a Pulse Per Second (PPS), enforcing a specific start time for any observation. The timestamp for any given portion of the dataset can be calculated relative to this starting time. Accordingly, no other synchronization or timing devices are required. 

\subsection{Software Packages}
\label{ss:ds}

\subsubsection{FPGA}
\label{ss:fpga}

The FPGA firmware was developed using Matlab/Simulink and Xilinx System Generator\footnote{http://www.xilinx.com/tools/sysgen.htm}, coupled with libraries developed by the Collaboration for Astronomy Signal Processing and Electronics Research (CASPER)\footnote{https://casper.berkeley.edu/wiki/Libraries}. These configured the signal routing and FPGA logic for  capture of samples from the ADCs, polyphase filter bank (PFB) operation, Fast Fourier Transform (FFT)  processing, and packetizing for ethernet transmission off the ROACH.    

\subsubsection{CPU}
\label{ss:cpu}

The data transmitted from the ROACH hardware are received using the Pulsar Distributed Acquisition and Data Analysis \texttt{PSRDADA}\footnote{http://psrdada.sourceforge.net} software package.  The core functionality of the package is the capture and subsequent streaming of data between ringbuffers. Data are acquired from a network interface or other direction connection and stored in a buffer.  Each buffer is broken up into a header block, containing information identifying the origin of the data and several sub-blocks. As data are received they are written to the sub-blocks sequentially. When a sub-block is full, a flag is raised signaling that the data in the sub-block can now be read. Read and write clients manage this process, such that data can be read from one buffer, manipulated, and written to another. This forms a modular processing chain into which additional steps can be inserted without modifying other aspects of the system. 

\subsubsection{GPU}
\label{ss:gpu}

The open \texttt{xGPU} package first developed by  \citet{clark2012}--a.k.a. the Harvard X-engine-- is an optimized scalable cross-multiplication code. The required performance of the CPU and GPU layers in combination is measured via the  rate of single precision (SP) floating point operations, R, and  capture rate, D (including all stages of the \texttt{PSRDADA} pipeline and GPU calculations).  Both depend on the number of correlator inputs  (N$_i$: twice the number of antennas or elements for two polarizations), number of frequency channels N$_c$, and the Nyquist sample interval for the F-engine spectra ($\tau$): 

\begin{equation}
\label{eqn:tflop}
\textrm{R}= 4 \textrm{N}_{i} ( \textrm{N}_i + 1) \textrm{N}_c \tau^{-1};
\end{equation}

\begin{equation}
\label{eqn:gbps}
\textrm{D} = 8 \textrm{N}_i \textrm{N}_c \tau^{-1}.
\end{equation}
On GPUs, there are four floating point operations per fused complex multiply and add operation.  Thus, dividing equation 1 by four enables crude comparison with the operations count for correlation using fixed-point arithmetic. 

\subsection{Frequency Domain Transformation}
\label{sec:dataacq}
\begin{figure}
\begin{center}
\includegraphics[width=7cm]{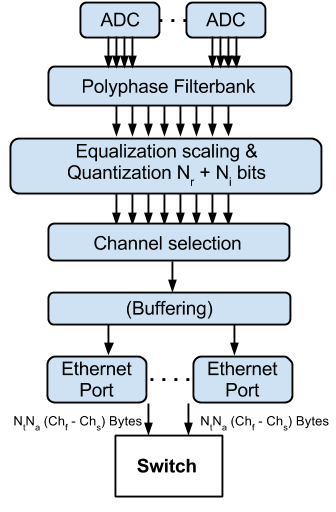}
\caption{FPGA processing pipeline. Digitized inputs are sent from the ADC to a PFB. Post PFB the signals are scaled to have 4-bits for the real, and 4-bits for the imaginary components. A selection of frequency channels are then buffered and transmitted via ethernet ports connected to the FPGA. Here, N$_t$ is the number of time samples per packet, N$_a$ the number of ADC inputs and Ch$_s$ and Ch$_f$ the starting and finishing frequency channels for the current packet. \label{fig:feng}}
\end{center}
\end{figure}

The correlator F stage comprises four operations: a PFB,  equalization and requantization, channel selection, and packet assembly (figure\,\ref{fig:feng}).  Each sampled data stream is channelized in frequency using a PFB.  The number of bits required to represent the data typically increases during this stage (e.g. from 8 bit (real) samples at the ADC to 36 bits (complex) samples after the PFB). Since the signal in each frequency channel of each Nyquist sample is noise-like, each channel is requantized to four bits for each component, real and imaginary \citep{backer2007, parsons2008}.  This reduces the F-stage output data rate, and network load.Frequency channel selection further reduces the network load by restricting later processing to frequency sub-bands that are free of persistent terrestrial interference.

The final stage before transmission of the data is the buffering of multiple spectra. As the number of inputs increases, the number of frequency channels that can be processed per GPU decreases, and frequency channels must be distributed among a larger number of servers. Buffering enables multiple Nyquist sampled spectra to be sent in a packet, keeping the packet size large when few frequency channels are sent per packet. This reduces the network overhead, and enables higher data rates than can be achieved with small packets. A 16-byte header is added to each packet for identification purposes. This header contains a packet sequence number (incremented with each packet) and an F-engine identification number. These two numbers allow the packets to be placed in the correct memory location when received, regardless of the order in which they arrive. These identification numbers take up only a fraction of the header space, allowing for additional information (e.g. information regarding RFI) to be encoded with the packet in the future (see figure \ref{fig:packet}).

\subsection{Corner Turn and Cross-multiplication}
\label{ss:ct}

\begin{figure}
\begin{center}
\includegraphics[width=8cm]{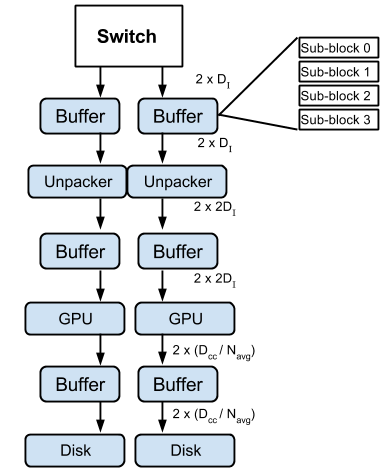}
\caption{Processing stages and buffers for a dual GPU receiving system. Each ``Buffer'' block in the diagram represents a ring buffer with four sub-blocks. As each sub-block is filled, the next stage in the pipeline takes control. Separate threads manage data capture, unpacking from 4 to 8 bits and reordering, and GPU communication.\label{fig:dataflow}}
\end{center}
\end{figure}

Data received from the F stage are processed through a series of ringbuffers (figure \ref{fig:dataflow}). The first stages in the pipeline, data capture and unpacking, complete the corner turn initiated in the network layer, and prepare the data for input to the GPUs for cross-multiplication. 

In the context of a correlator, the corner turn operation is the process of assembling frequency channels from multiple inputs, for example, frequency channel zero from every input antenna polarisation. When all inputs are handled by a single device, this process is the same as a matrix transpose (e.g. transposing all the parallel inputs so they are ordered by frequency channel rather than input). When the inputs originate from multiple devices, there must be an interconnect to allow the frequency channels from different F-engines to be merged. In a similar manner to that described in \citet{parsons2008}, a switch is used to simplify implementation of the corner turn operation in this design. Instantiation of the corner turn in the network and CPU layers simplifies and streamlines the operation, playing a critical role in enabling scalability to at least O(10$^3$) inputs without design alteration (see Section \ref{sec:scale}).

\begin{figure}
\begin{center}
\includegraphics[width=8cm]{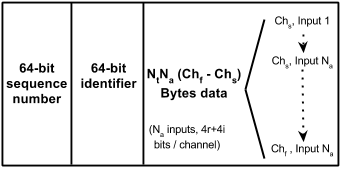}
\caption{Packet format. The 64-bit sequence number combined with the F-engine identifier creates a unique identifier for each packet received by the server. The combination of these two numbers also indicate where in memory the packet should be placed. The packet size is dictated by the number of frequency channels each GPU can process. Where multiple Nyquist sampled spectra are sent in a single packet, each new time sample, formatted as per the data portion of the packet, is added to the end. \label{fig:packet}}
\end{center}
\end{figure}

Once captured over an ethernet interface, packets are placed into RAM buffers based on the identification number of the packet and the originating F-engine, listed in the header (figure \ref{fig:packet}). If a buffer boundary is reached before all the data for the buffer has been received due to out of order packets, packets for the next buffer can be stored in stack memory temporarily. Once the number of packets in the stack reaches a pre-defined threshold, the remaining missing packets are recorded lost, and the buffer marked as full. The total data rate to be captured by each server, D$_I$, is given by
\begin{equation}
\label{eqn:datain}
\textrm{D}_I = [8 \textrm{N}_r \textrm{N}_a (\textrm{High}_{ch} - \textrm{Low}_{ch}) ] / [\textrm{N}_{serv} \tau]
\end{equation}
where N$_r$ is the number of FPGA nodes, N$_a$ is the number of inputs per FPGA node, Low$_{ch}$ and High$_{ch}$ are the starting and finishing channel numbers of the selected frequency band respectively, and N$_{serv}$ is the number of servers.

\begin{figure}
\begin{center}
\includegraphics[width=7cm]{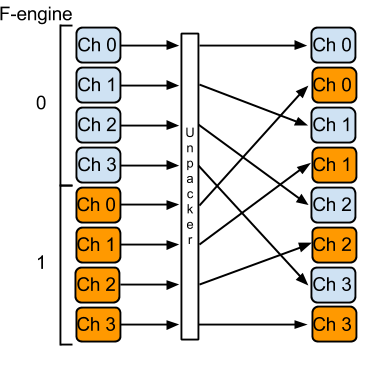}
\caption{Data flow during unpacking. The unpacker in the CPU layer completes the corner turn by reordering the frequency channels received from each FPGA to be adjacent. The blue and red blocks indicate frequency channels from different F-engines and antenna inputs.\label{fig:unpacker}}
\end{center}
\end{figure}

The capture of data and placement in RAM buffers completes the first part of the corner turn: the data packetized on the FPGA are transmitted to the appropriate GPU, coalescing the frequency channels for each subset of bandwidth. After capture, a separate process pads the data from four to eight bits. While frequency channels from each FPGA node will be grouped together (e.g. channel zero from each ADC input), the channels from different nodes will not be contiguous in memory. Therefore, in addition to bit promotion, the ``unpacking'' step in the dataflow includes an implicit reordering within the defined range assigned to a GPU node and thereby completes the corner turn (figure \ref{fig:unpacker}). Once the data are 8-bit, they can be passed through the GPU texture memory.  The texture memory converts the input from 8 to 32 bits in hardware, effectively making the 32 bit promotion ``free'' in terms of CPU/GPU cycles \citep{clark2012}. The data are then cross-multiplied in the GPU as in \citet{clark2012}. The data size of the cross-correlation matrix to be output from each GPU, D$_{cc}$, is calculated by:

\begin{equation}
\textrm{D}_{cc} = \textrm{N}_c (\textrm{N}_i / 4 + 1)(\textrm{N}_{i}/2) \textrm{N}_{pol}^2. 
\end{equation}
where N$_{pol}$ is the number of polarizations. In order to make the output data rate manageable, this matrix is averaged on the GPU before transferring back to host memory, and can then be further averaged on the CPU before writing to disk. 
 
\section{Scalability}
\label{sec:scale}

While the requirements for scalability of a signal processing system are superficially rudimentary (e.g., increase processing power and memory size), in practice, it cannot be taken as given that any particular architecture can be scaled to larger N when details of implementation are considered. The modular nature of the hybrid correlator, choice of packet format and corner turn implementation, enable scaling as a function of inputs and bandwidth as described by equations \ref{eqn:tflop}  - \ref{eqn:datain}. As a limit is reached in the number of inputs, GPU processing capacity or data capture rate, additional F-engine nodes may be added to enable new signal paths, and X-engine nodes to expand cross-multiplication capacity. Note that in order for the corner turn to scale, the requisite number of switch ports increases concomitantly and a full cross-bar network configuration is needed when the number of required ports exceeds the capacity that can be accommodated by a single switch\footnote{This could also be accomplished using multiple smaller parallel networks.}.

In addition to scaling the number of inputs, the F-stage can similarly be scaled in bandwidth using standard techniques such as using multiple oscillators to split the input baseband between platforms, or constructing a compound F-engine (e.g., a course PFB followed by a fine PFB). Scaling of the F-engine may be limited by packet buffering. As the number of antennas grows, increasing numbers of Nyquist sampled spectra need to be buffered on the FPGA before transmission. This can be accomplished using the FPGA buffer random access memory (BRAM), or if required, additionally attached memory such as quad data rate (QDR) or dynamic random-access memory (DRAM). The packet size can be calculated by N$_t$N$_a$N$_c$, where N$_t$ is the number of time samples per packet. The number of packets that can be buffered is then M$_{size}$ / N$_t$N$_a$N$_c$, where M$_{size}$ is the amount of buffer memory available. Taking the ROACH2 platform as an example, there is 36~MB of QDR memory. This allows for buffering of order 4400, 8$\,$kB packets, through to 35000, 1$\,$kB packets, where the minimum packet size will be limited by the required data capture speed for each GPU server. 

There are two points at which GPUs may limit scalability. The first occurs when the processing requirements of the system are such that a GPU cannot process a single frequency channel in real time. The second case is a hardware limit, where the number of inputs increases to where the data for a single frequency channel will not fit on the device. This first can be overcome within the current architecture by assigning multiple GPUs to the same frequency channel, and multiplexing in time. For example, GPU zero processing frequency channel zero at time step one, and time step three, while GPU one processes time steps two and four. An alternative option involves partitioning the visibility matrix so that each GPU computes and stores only a subset of the matrix. This can be thought of as an additional level of memory tiling for the GPU. This partitioning solves both limitations simultaneously. Using current generation hardware, the maximum scaling is to $\sim$5600 antennas without time multiplexing or partitioning assuming a sustained processing rate of 3~TFlop\,s$^{-1}$ per GPU. Assuming a $\sim$6~GB memory limit, the maximum number without partitioning is $\sim$19000. 

Table \ref{table:datarate} shows capture rates\footnote{Prior to unpacking.} and Tflop\,s$^{-1}$ per GPU for a 32, 64, 512, 5600 and 19000-input system, as well as Top\,s$^{-1}$ for the F-engine. For the lower inputs systems, it is data rate that is the dominating factor. For larger numbers of antennas the data rates into each GPU can become negligible. The 19000-input system assumes that multiple GPUs are assigned to each channel. 

\begin{table*}
\begin{center}
\caption{\rm Data rate and processing requirements for a varying number of inputs.\label{table:datarate}}
\begin{tabular}[t]{c|c|c|c|c|c}
\hline
\hline
Inputs                      & 32          & 64        & 512    & 5600  & 19000  \\ 
\hline
FPGA processing &&&&&\\ 
(Top\,s$^{-1}$ required)    & 0.01        & 0.02      & 1.81   & 1.98  & 6.72 \\
\hline
\hline
GPU processing &&&&&\\
(TFlop\,s$^{-1}$ required)  & 0.42        & 1.66      & 105    & 12539 &  144324 \\ 

\hline
GPUs (3TFlop\,s$^{-1}$)              & 1           & 1        & 36     & 4180  &  48109   \\ 
GPUs (24TFlop\,s$^{-1}$)             & 1           & 1        & 5      & 523   &  602     \\ 
\hline
Data (Gbps) / GPU (3TFlop\,s$^{-1}$)    & 25.58       & 51.17    & 11.37   & 1.07  & 0.316\\ 
Data (Gbps) / GPU (24TFlop\,s$^{-1}$)     & 25.58       & 51.17    & 81.88   & 8.56  & 2.53 \\ 
\hline
\end{tabular}
\end{center}
\end{table*}

\begin{figure}
\begin{center}
\includegraphics[width=9cm]{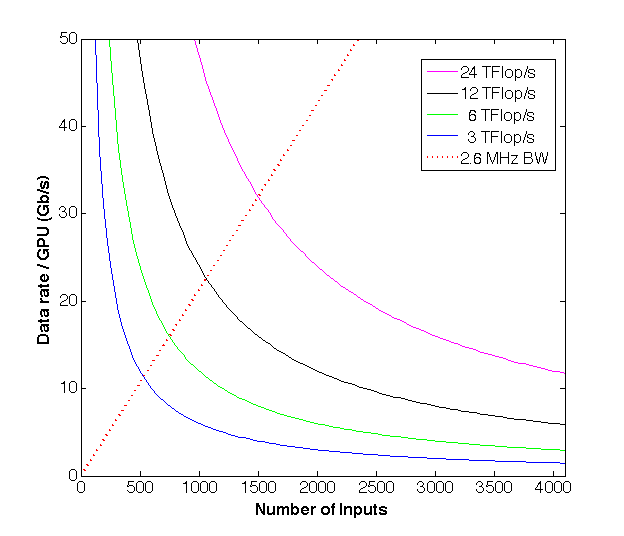}
\caption{Computational optimization as a function of input count. Solid curves represent the trade off between the number of inputs and bandwidth that can be processed, for GPU single precision floating point capacities of 3, 6, 12, and 24~TFlop\,s$^{-1}$ (actual).  The red dotted line specifies the data rate for the specific example of a 2.6~MHz bandwidth.  Above the line processing is bandwidth bound. At the intersection of the lines, the system becomes computationally bound. Below the line the data are unable to be processed in real time.\label{fig:scale}}
\end{center}
\end{figure}

An ``ideal'' system for hardware utilization maximizes both both throughput and computation. Figure \ref{fig:scale} illustrates the continuum of the Table \ref{table:datarate} estimates, visualizing the potential tradeoffs that can be made. For a GPU with 3~TFlop\,s$^{-1}$ sustained processing capacity for example, correlating 2.6~MHz of bandwidth, an optimal system has $\sim$500 inputs. For a 24~TFlop\,s$^{-1}$ capable GPU, the ideal number of inputs increases to ~$\sim$1500. For fewer inputs, the GPUs are not computationally bound. In this case GPUs with a lower peak performance can be used to reduce costs. For greater numbers of inputs the GPUs are not capable of processing the requested bandwidth in real time. 

The scalability aspects discussed here are focused solely on the correlator. While this may be sufficient for some applications, in the general case other elements in the data path pose unsolved problems of scale owing to dependence on at least N$^2$. The computation challenges associated with gridding irregularly spaced visibilities in preparation for FFT imaging \cite{romein2012}, and subtraction of sky models from correlator output in the visibility domain \cite{mitchell2008}, for example, will also need to be addressed.

\section{Specific Implementations \& Field Demonstration}
\label{sec:results}

\begin{figure}
\begin{center}
\includegraphics[width=7cm]{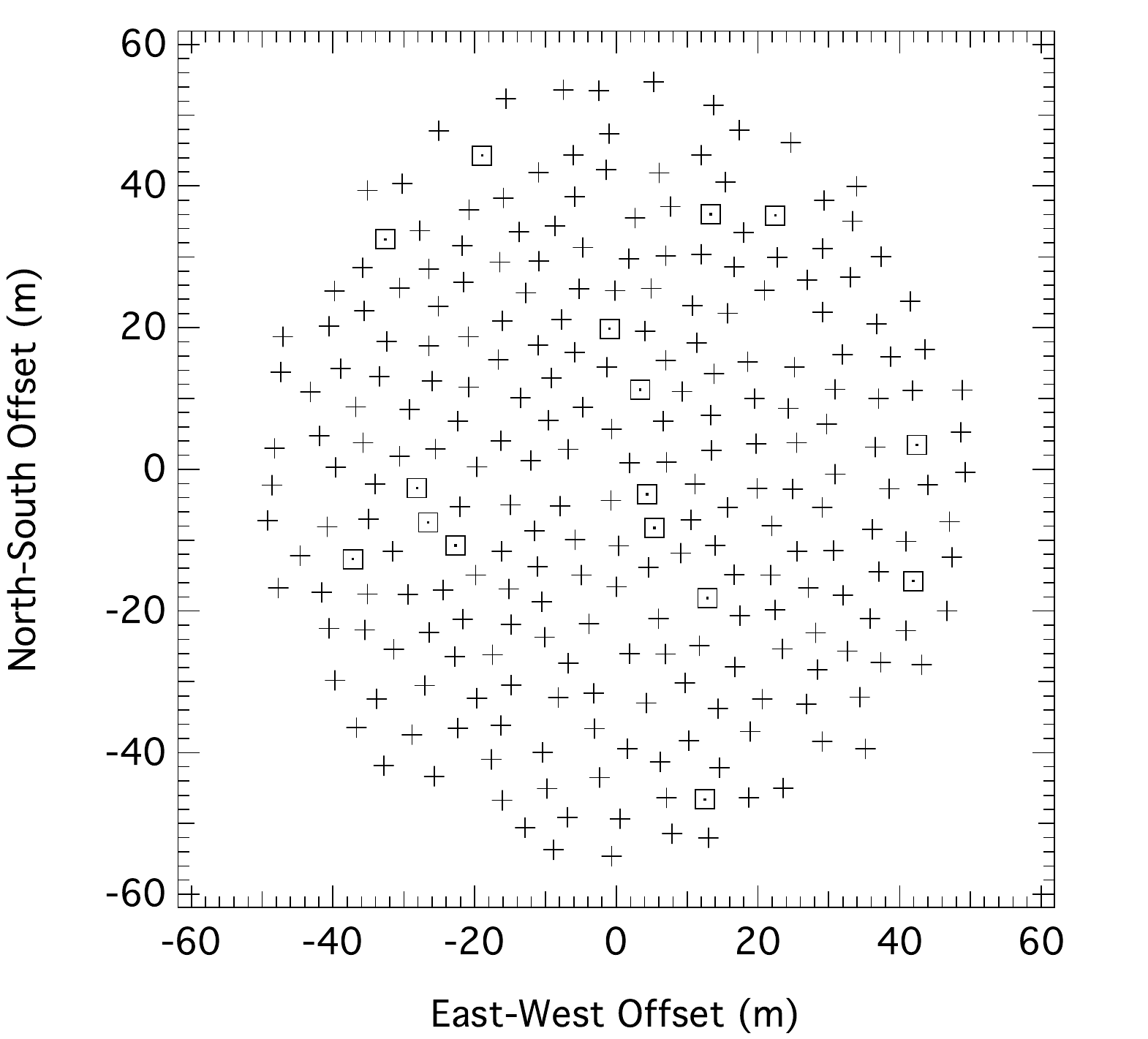}
\caption{LWA1 stands layout. The stands used for the 32-input demonstration are represented by squares.\label{fig:stands}}
\end{center}
\end{figure}

Operation of the hybrid correlator configured for 32 inputs was initially demonstrated using a subset of antennas at the LWA1 site in New Mexico \citep{taylor2012, ellingson2013}. The subset comprised 16 dual polarization dipoles at spacings from $\sim$5 to 98~m (figure \ref{fig:stands}).

Four ROACH boards, housing dual, quad input, ADCs were used to digitize the input signals\footnote{https://casper.berkeley.edu/wiki/ADC4x250-8}. The ADCs were clocked at 200~MHz, to achieve a 100~MHz bandwidth. A PFB (2-tap, 8192~pt resulting in 24.4~kHz channel spacing), transformed the data to the frequency domain (figure \ref{fig:feng}). The data were quantized to 4-bits real and 4-bits imaginary for each frequency channel, and a contiguous frequency sub-band of $\sim$40~MHz was selected from the output (1628 channels).

Frequencies above 88~MHz are dominated by FM radio broadcasting.  Below 30~MHz at the LWA1 site, the time and frequency occupancy of interference increases toward the low end of the band, principally due to long range propagation conditions created by the ionosphere.\footnote{See http://www.fcc.gov/oet/spectrum/table/fcctable.pdf for the spectrum allocation.} Above 54~MHz, television broadcasting may be anticipated in general, but the LWA1 site is primarily clear of persistent local sources due to the transition to digital TV that moved most of the broadcasting to frequencies above 100~MHz.

Each sub-band was formatted into 10~GbE and transmitted via a switch to a GPU server\footnote{Dual E5645 CPU, dual Tesla C2050 GPUs, dual 10GbE ports}. The 32-input design required that data be captured into RAM at an input rate of $\sim$10~Gbps. The data were unpacked and cross-multiplied in GPUs, per figure \ref{fig:dataflow}. For this implementation, one CPU thread was used to capture on each 10~GbE port (the input data were split between two ports running at approximately 5~Gbps each), two threads were used for each unpacker and reorder step, one thread to call the GPU kernel, and one thread to average and write the data output to disk\footnote{Tests have shown with current generation hardware the software used is capable of capturing and processing data at 9.9~Gbps for each 10~GbE link, or alternatively up to 16~Gbps for each CPU thread when using a 40~GbE link, and that the unpacker code can process approximately 15~Gbps per CPU thread. Data capture and processing using multiple threads has shown the pipeline capable of capturing 39.6~Gbps and processing 30~Gbps of input data. The GPU performance is analyzed in \cite{clark2012}}. Each GPU pre-averaged 1024 cross-spectra, and the CPU a further 119, resulting in output time averaging of $\sim$4.99 seconds. 

\begin{figure}
\begin{center}
\includegraphics[width=7cm]{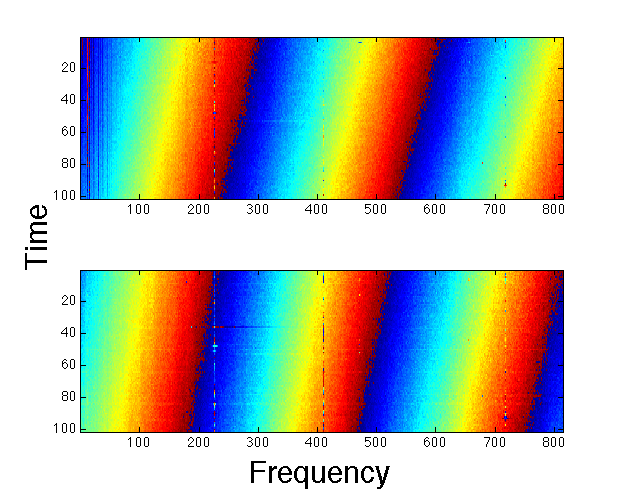}
\caption{Fringes for a 91~m baseline, in two orthogonal polarizations. The vertical axis shows 5 second time steps.,The horizontal axis frequency steps 24~KHz channels. The phase is coded by color (red-blue is one turn).  The two LWA1 stands used for the figure formed a baseline 29.7 degrees west of north.\label{fig:fringes}}
\end{center}
\end{figure}

\begin{figure}
\begin{center}
\includegraphics[width=7cm]{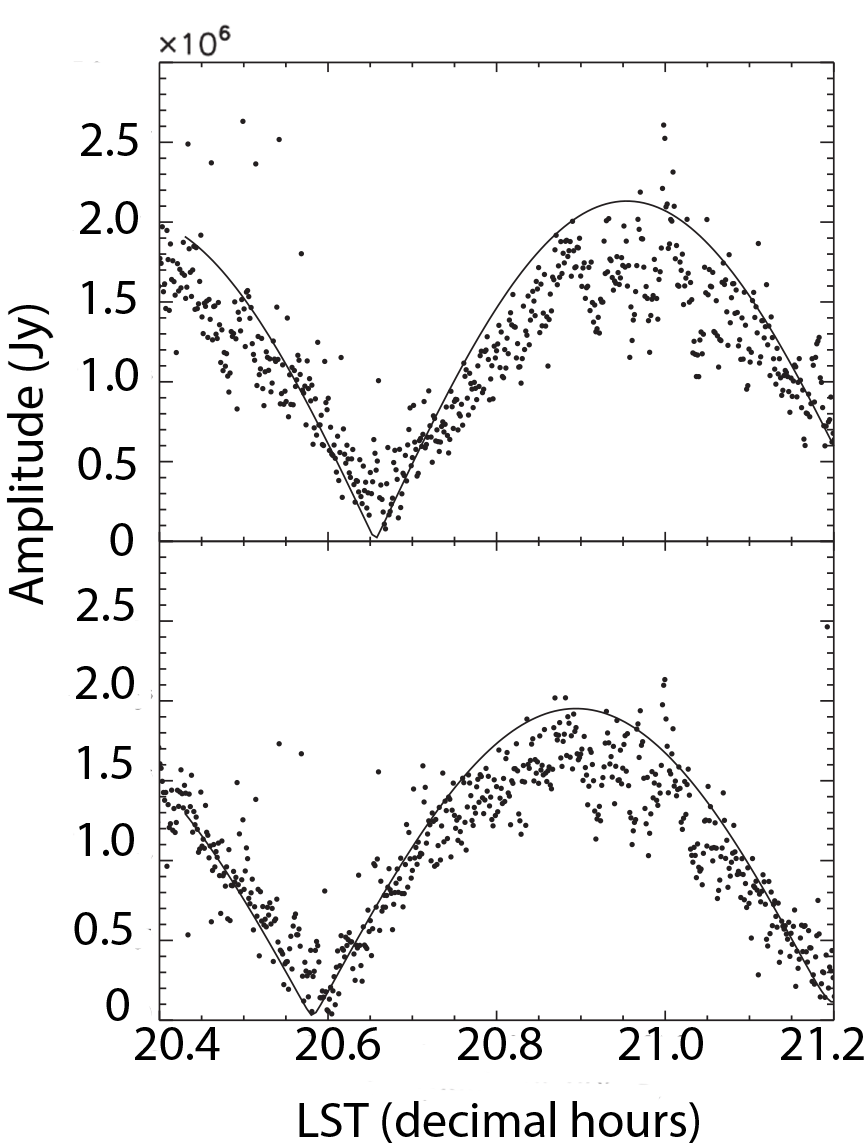}
\caption{Data from observations dominated by Cygnus A and Cassiopeia A for two different baselines. Visibility amplitudes as a function of time for a 91~m baseline at a position angle of 29.7 (top panel) and 98~m at 18.7 (bottom panel) west of north. These baselines substantially resolve out diffuse galactic emission. The visibility amplitude expected for a sky model of two point sources, representing CasA and CygA (solid line) is scaled to match the amplitude of the observed data (dots). The data are shown for one polarization, and the amplitude is an average over three 24.4~kHz frequency channels at 52 MHz. The fringe patterns show ``beating'' between the two sources. \label{fig:ampvtime}}
\end{center}
\end{figure}

\begin{figure}
\begin{center}
\includegraphics[width=7cm]{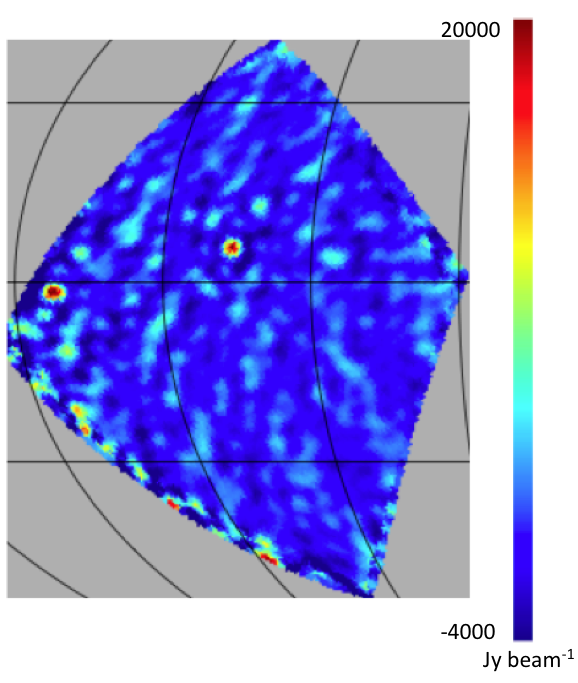}
\caption{Cygnus A and Cassiopeia A in an instantaneous 90 degree field of view. CygA is used as the Stokes I calibrator, adopting flux densities of 17000 Jy at 74 MHz \citep{cohen2007} and 22000 Jy at 38 MHz \citep{kellermann1966}. The flux and position of Cassiopeia A are within 4\% and 0.7\% of those in the literature respectively \citep{kellermann1966}. Data analysis was carried out as in \cite{bernardi2013}. After bandpass calibration and fit for the dipole complex gains as a function of time \citep[as described in][]{mitchell2008}, the visibilities were Fourier transformed to snapshot images that were mosaiced together \citep{ord2010}. CygA and CasA were deconvolved using the forward modeling technique described in \cite{bernardi2011}.\label{fig:image}}
\end{center}
\end{figure}

A selection of fringes obtained following the 32-input installation can be seen in figure \ref{fig:fringes}, and figure \ref{fig:ampvtime} compares the progression in time of the fringe amplitudes at 52~MHz to a two source model for different baselines. Finally, an image of Cygnus A and Cassiopeia A based on 5 minutes of data with 20~MHz bandwidth is given in figure \ref{fig:image}.

For the GPUs used in the 32-input test system, the \texttt{xGPU} algorithm can achieve up to 79\% of the GPU peak performance \cite{clark2012}. For the test system, the number of inputs was trivially small, needing less than 10\% of the available resources to compute the cross-multiplication (0.06~Top\,s$^{-1}$). The F-engine was divided over four FPGAs, each processing eight inputs. The total computation for the F-stage was correspondingly small (0.1~Top\,s$^{-1}$), requiring each FPGA to process 0.025~Top\,s$^{-1}$. \footnote{Approximately 7\% of the total theoretical available on a Virtex 5 SX95T chip assuming the maximum clock rate of 550~MHz could be achieved and all DSP48E slices were used. Approximately 20\% when the actual clock rate of 200~MHz is used.} Larger 64 and 512-input systems that more fully utilize the corresponding hardware \citep[e.g. as in][]{clark2012} have been employed. Each system processes $\sim$60~MHz of bandwidth. The 64-input system replaced the original 32-input at LWA1, and the 512-input correlator was installed at the LWA station at Owens Valley Radio Observatory (LWA-OVRO).

\section{Summary \& Future Work}
\label{sec:conclusions}

The ``hybrid'' correlator is a scalable large-N correlator implemented in both hardware (FPGA) and software (CPU/GPU). The design architecture described has been successfully implemented at LWA1 as a 32 and 64-input system, and at LWA-OVRO with 512 inputs. Without modification, the design should be scalable to past the O(10$^3$) antenna regime.  Planned developments for the correlator pipeline include baseline dependent integration (BDI), pulsar gating and partitioning of the visibility matrix. Shorter baselines will have slower fringe rotation, and can be integrated longer without decorrelation. BDI will enable the integration length for different baselines to be specified, decreasing the output data rate. Pulsar gating will allow placing the ``on'' and ``off'' pulse time samples for pulsars over the full field of view into separate data streams, enabling generalization of the calibration scheme discussed in \cite{pen2009}. Partitioning of the visibility matrix will also be implemented to overcome potential limitations in the GPU hardware when dealing with O(10$^4$) antennas.

\section*{Acknowledgments}

Research presented here was supported by National Science Foundation grants PHY-083057, AST-1106045, AST-1105949, AST-1106059 and AST-1106054.
The authors acknowledge contribution from the Long Wavelength Array facility in New Mexico, which is supported by the University Radio Observatories program under grant AST-1139974, and National Science Foundation grant AST-1139963.

{\it Facilities:} LWA

\end{document}